\documentclass[12pt]{iopart}
\usepackage{iopams}  
\usepackage{hyperref}
\usepackage{graphicx}

\begin{document}

\title{Evaluation of ambipolar potential barrier in the gas dynamic trap by Doppler spectroscopy}

\author{$Andrej~Lizunov^{\dag}, Vladimir~Maximov^{\dag}, Andrey~Sandomirsky^{\ddag}$}
\address{\dag Budker Institute of nuclear physics, 630090 Novosibirsk, Russia} 
\address{\ddag Novosibirsk State University, 630090 Novosibirsk, Russia} 
\ead{lizunov@inp.nsk.su}
\vspace{10pt}

\begin{abstract}
The recently developed Doppler spectroscopy diagnostic has been used to evaluate the height of the ambipolar potential barrier forming in the gas dynamic trap (GDT) plasma between the central cell and the region with a large magnetic expansion ratio beyond the mirror. The diagnostic technique based on the gas jet charge exchange target, allowed to measure the potential profile along the line of sight covering the radial range from the axis to the limiter. The on-axis potential drop was found to be $2.6 \div 3.1$ in units of the central plane electron temperature, which supports the existing theoretical understanding of suppression of electron thermal conductivity in the GDT expander.
\end{abstract}

%
\vspace{2pc}
\submitto{\NF}
%
\maketitle
%
%

\section{Introduction}
\label{intro}
\begin{figure}[htbp]
\centering 
\includegraphics[width=0.7\textwidth]{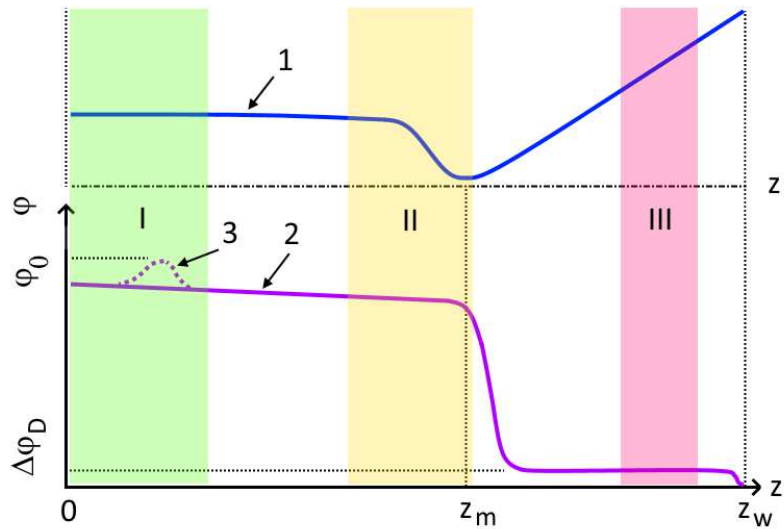}
\caption{\label{potential} Axial profile of magnetic field line and ambipolar potential in GDT: 1 -- magnetic field line, 2 -- potential profile peaked in the centre, 3 -- potential profile peaked in the fast ion turning point. I -- central zone, II -- mirror zone, III -- location of CXRS diagnostic gas target.}
\end{figure}

Linear magnetic mirror traps for plasma confinement have an intrinsically simpler design comparing to toroidal systems. This simplicity is the most ample for an axially symmetric magnetic field configuration. The gas dynamic trap (GDT) \cite{gdt-review-ppcf} is a device of this kind for confinement of two-component plasmas with anisotropic fast ions generated with angled injection of eight 25~keV deuterium beams firing the total power of $\approx 5$~MW. In a classical scenario, the ``target'' plasma remains in a strongly collisional confinement regime. That means that the axial particle outflow is proportional to the density and the mirror ratio (which is relatively large, $k_m = H_m / H_0 \geq 40)$. A similar gas dynamic flow scaling would be inherent for the pressurised vessel with a pinhole leak. Numerous direct measurements in GDT demonstrate the peak transverse plasma $\beta$ reaching 0.6 \cite{gdt_maxbeta} within a pulsed discharge with the heating stage duration of only $\approx 10$~ms. The later upgrade with the ECRH hardware has allowed to extend the range of available electron temperatures in confined plasmas beyond one keV \cite{ecrh_prl}. These results among other recent achievements in the gas dynamic trap \cite{cool_mirror_results} could not be real without a strongly pronounced suppression of axial heat flux through the mirror with respect to a straight-forward exchange of hot and cold electrons on the end wall \cite{spitzer}. It is a common knowledge that any competitive concept of a linear magnetic trap for fusion (see, for example \cite{gdt_ns-1, gdt_dia_trap-1, gdmt, tae_Gota_2017}) would ultimately require a radical depression of axial heat losses. The basic physics of axial plasma confinement in the gas dynamic trap with the magnetic expander section has been reviewed in \cite{axial_conf}. The theory shows the longitudinal profile of the plasma electrostatic potential with the maximum at the centre of the confinement section and the major slope located beyond the mirror waist as schematically drawn in Fig.~\ref{potential}. The potential shape in the central zone follows either curve 2 or 3 depending on the fast ion density in the turning point relatively to that of the bulk plasma. In either case, we mark $\varphi_0$ the maximum of potential in the central zone. For given experiments with the typical target plasma density of $n_p\leq 2\cdot10^{19}~m^{-3}$ and the ion temperature of $\approx 100$~eV, the warm ion scattering mean free path is $\lambda_i \geq 5$~m which notably exceeds the half mirror-to-mirror distance of 3.5~m. A rather moderate temperature of $100\div250$~eV characterises the electron component as no auxiliary ECR heating is used in the campaign. Under these conditions, the bulk ions containment is a some transient mode between the gas dynamic and adiabatic.  One should treat this task rather in terms of numerical model than analytically. However, for the sake of a simpler interpretation of observed data, one can depart from the collisionless equations for the energy and magnetic moment conservation:
\begin{equation}
\label{ion_energy}
\frac{m_i v^2}{2} = \frac{m_i v_0^2}{2} + \Delta U(z),
\qquad
 v_{\perp}^2 = v_{\perp_0}^2 \frac{H(z)}{H_0},
\end{equation} 
where $U(z) = e(\varphi_0 - \varphi(z))$. The ambipolar potential profile plotted as curve 2 and 3 in Fig.~\ref{potential}, forms a barrier repelling the majority of hot electrons back to the central confinement section from the mirror. Its height develops due to the electron distribution functions (EDF) in the central cell and expander considering the conditions of quasineutrality and equality of current densities of electrons and ions \cite{axial_conf}. Even for the secondary emission coefficient close to unity on the end wall, the backward flow of cold electrons through the mirror is effectively depleted due to the magnetic moment conservation if the magnetic expansion ratio is large enough on the wall: ${k_w}^{-1} = H_0 / H_w \geq \sqrt{m_i / m_e} \approx 63$ \cite{axial_conf, gdt-review-ppcf}. Buildup of the trapped electron fraction in the expander EDF leads to flattening out the potential distribution as schematically illustrated in Fig.~\ref{potential} in the region $z > z_m$. Assuming this model, the potential does not vary significantly in the expander along the field line up to the Debye sheath on the wall encapsulating the remaining drop $\Delta \varphi_D \simeq T_e$.

\section{Measurement of potential by Doppler spectroscopy}
\label{measurement}
\begin{figure}[htbp]
\centering 
\includegraphics[width=0.9\textwidth]{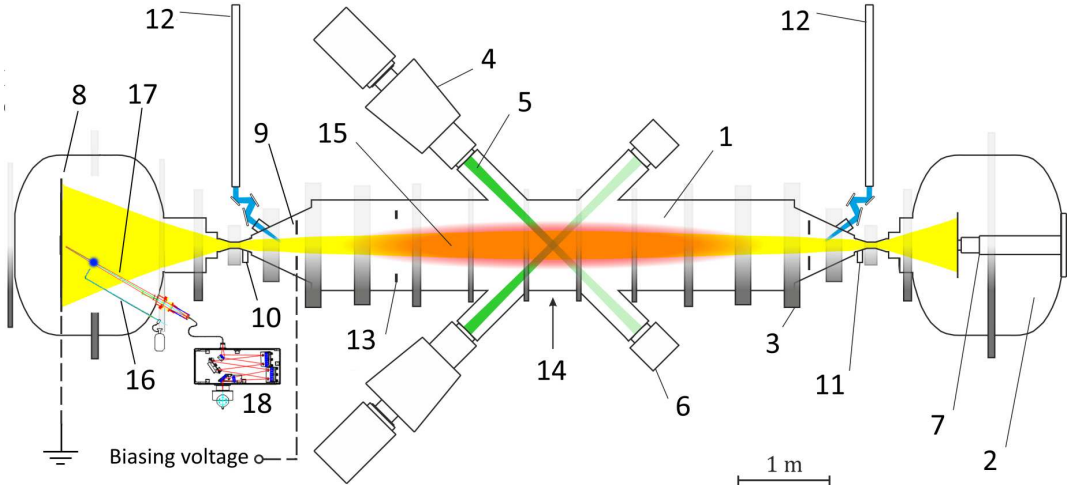}
\caption{\label{gdt} The gas dynamic trap: 1 -- central cell, 2 -- right expander tank, 3 -- magnetic coil of central solenoid, 4 -- atomic beam injector, 5 -- deuterium beam, 6 -- beam dump, 7 -- arc discharge plasma source, 8 -- plasma dump in the left expander tank, 9 -- radial limiter, 10 -- left gas box, 11 -- right gas box,  12 -- waveguides of ECRH system, 13 -- diamagnetic loop, 14 -- Thomson scattering diagnostic, 15 -- fast ions, 16 -- movable gas puff tube of CX target, 17 -- CXRS line of sight, 18 -- spectrometer with CCD camera.}
\end{figure}

The GDT layout is shown in Fig.~\ref{gdt}, several main parameters of the machine and the experimental scenario are listed in the Table~\ref{gdt_param}.

\begin{table}[htbp]
\caption{\label{gdt_param} The main GDT parameters.}
\footnotesize
\begin{tabular}{@{}llll}
\br
\it{Parameter} & \it{Value} \\
\mr
Mirror-to-mirror length & 7 m \\
Magnetic field $H_0$ & 0.35 T \\
Deuterium beam energy & 25 keV \\
Total NB power & 5 MW \\
NB pulse duration & 5 ms \\
Electron temperature & $100\div250$~eV \\
Fast ion mean energy & 10 keV \\
Plasma density & $1.5\cdot 10^{19} m^{-3}$ \\
Maximum $\beta_{\perp}$ & 0.5 \\
\br
\end{tabular}\\
\end{table}
\normalsize 

Though the energy balance patterns in the gas dynamic trap are well established at least for certainly collisional bulk plasmas, the properties of the ambipolar potential have not been studied exhaustively so far. Earlier experiments (see \cite{gdt-review-ppcf}, page 27) with the wall-mounted grid ion energy analyser have provided only on-axis data taken within a limited expansion ratio numbers. Multiple accompanying probe measurements typically show large discrepancies and generally are not trustworthy. Recently the approach of optical Doppler spectroscopy was successfully adopted to measure the potential drop and the ion temperature in the plasma flow in the GDT expander \cite{ELVIS}. The charge exchange radiation spectroscopy (CXRS) diagnostic currently has the single line of sight (LOS) hitting the machine axis at the plasma absorber plate as shown in Fig.~\ref{gdt}. The pulsed gas stream of $H_2$ puffed from the thin quartz capillary ({\it 16} in Fig.~\ref{gdt}) provides the target to produce charge exchange (CX) atoms emitting an optical signal of the good intensity for the spectral analysis. We observed a high contrast ratio $R_c =  Active~CX~signal / Background~signal \simeq 20$ in the working spectral regions comparing the spectra acquired with the CX gas target and without it. Within this accuracy, we can consider measurements local neglecting the line-integrated background light pickup. The localisation is defined by the CX target size $\Delta r_0 = \Delta r_t(z)\sqrt{H(z) / H_0} \leq 0.5~cm$. Here $\Delta r_t(z)$ and $\Delta r_0$ are transverse sizes in the measurement point and in the projection to the central plane along the magnetic surface, respectively. The gas tube {\it 16} is installed via the vacuum feedthrough allowing to translate it along the LOS, see Fig.~\ref{gdt}. Profiles of potential along the LOS are acquired repositioning the gas target from the plasma edge to the axis. Presuming a flat profile over $Z$ in the area {\it III} in Fig.~\ref{potential}, the transverse profile is similar to the profile over LOS.

Ongoing research programmes in GDT lean on the scenarios where the target plasma and heating beams are both deuterium. The plasma startup is initiated by the electron cyclotron discharge with a microwave burst of a moderate power \cite{ECR_startup}. Later on, the bulk plasma density is maintained with the peripheral puff of $D_2$ from the gas box integrated within the right mirror coil assembly, see Fig.~\ref{gdt}. For the given viewing geometry, Doppler spectroscopy measurements on deuterium are severely complicated by cold hydrogen emission which inevitably presents in GDT expander spectra. The noted inconvenience can be easily bypassed using hydrogen as a diagnostic impurity in the central cell. For that, the similar left-side gas box injects this impurity of $H_2$ into the plasma. This arrangement bounds the birth place of tracer $H^+$ ions to the vicinity of the left mirror, which is on the side of the spectroscopic diagnostic. Considering gas spreading over the plasma periphery, the upper estimate for the ionisation zone length might be one meter or so. This location is shown as the zone {\it II} in Fig.~\ref{potential}. The spectroscopic method yields an additional benefit like observation of multiple emission lines at once coming from different impurities. For that, the second tracer impurity of $He$ is injected in the central zone {\it I}, see Fig.~\ref{potential}. As it is described in \cite{ELVIS}, the monochromator is tuned in the spectral region for the simultaneous acquisition of two lines, $H_{\alpha}$ and $He$-$I$ with the wavelengths of $656.3~nm$ and $667.8~nm$ respectively. The accelerating potential drop extracted from the $H_{\alpha}$ Doppler shift is averaged over the birth place of $H^+$ ions -- the mirror zone {\it II}. In the same way, the  $He$-$I$ Doppler shift is linked with the averaged potential over the central zone {\it I}.

\section{Spatial profiles of potential and scaling over electron temperature}
\label{results}
\begin{figure}[htbp]
\centering 
\includegraphics[width=0.7\textwidth]{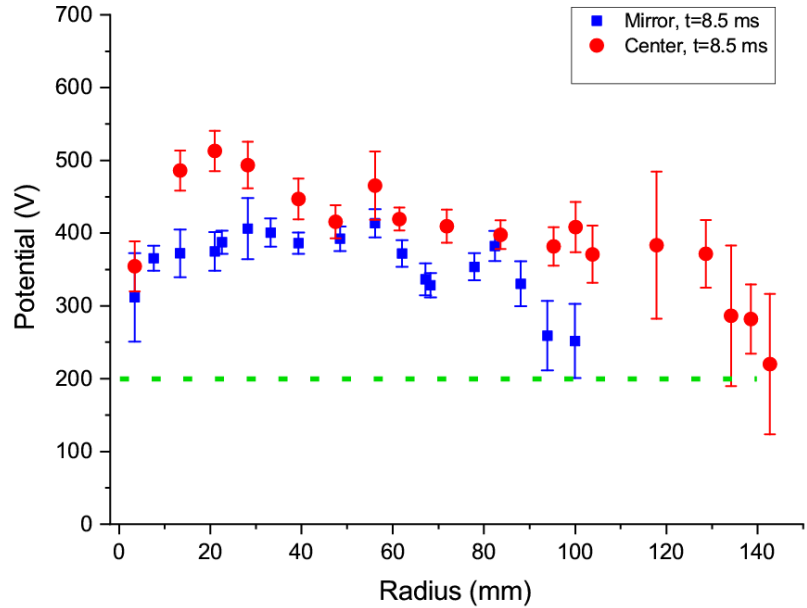}
\caption{\label{rad_profile} Radial profile of potential in the centre and in the mirror in the GDT central plane coordinates. Green dotted line -- bias voltage applied to the radial limiter at $r_0$=140~mm.}
\end{figure}

Independent scans over radial and axial coordinates would be necessary to recover the spatial distribution of the electrostatic potential drop. As Fig.~\ref{gdt} illustrates, both $r$ and $z$ coordinates change along the diagnostic line of sight so the scan across the LOS gives some slice of the spatial profile. This dependence is presented in Fig.~\ref{rad_profile} as a function of the radii along the LOS considering this note. Each point acquired over five to ten shots, where the error bars reflect the standard deviation. Each shot in a series delivered two data samples, one for the central potential and another one for the mirror.
As previously, the radial coordinate is projected to the GDT central plane along the magnetic surface. In measurements of the mirror potential, no valid samples were taken beyond $r_0 \simeq 100~mm$ due to the abrupt signal decrease. The profiles feature a small descent towards the axis, which is a more significant on the central potential. One may also notice the dip on the mirror potential profile around $r_0 \simeq 70~mm$. 

\begin{figure}[htbp]
\centering 
\includegraphics[width=0.9\textwidth]{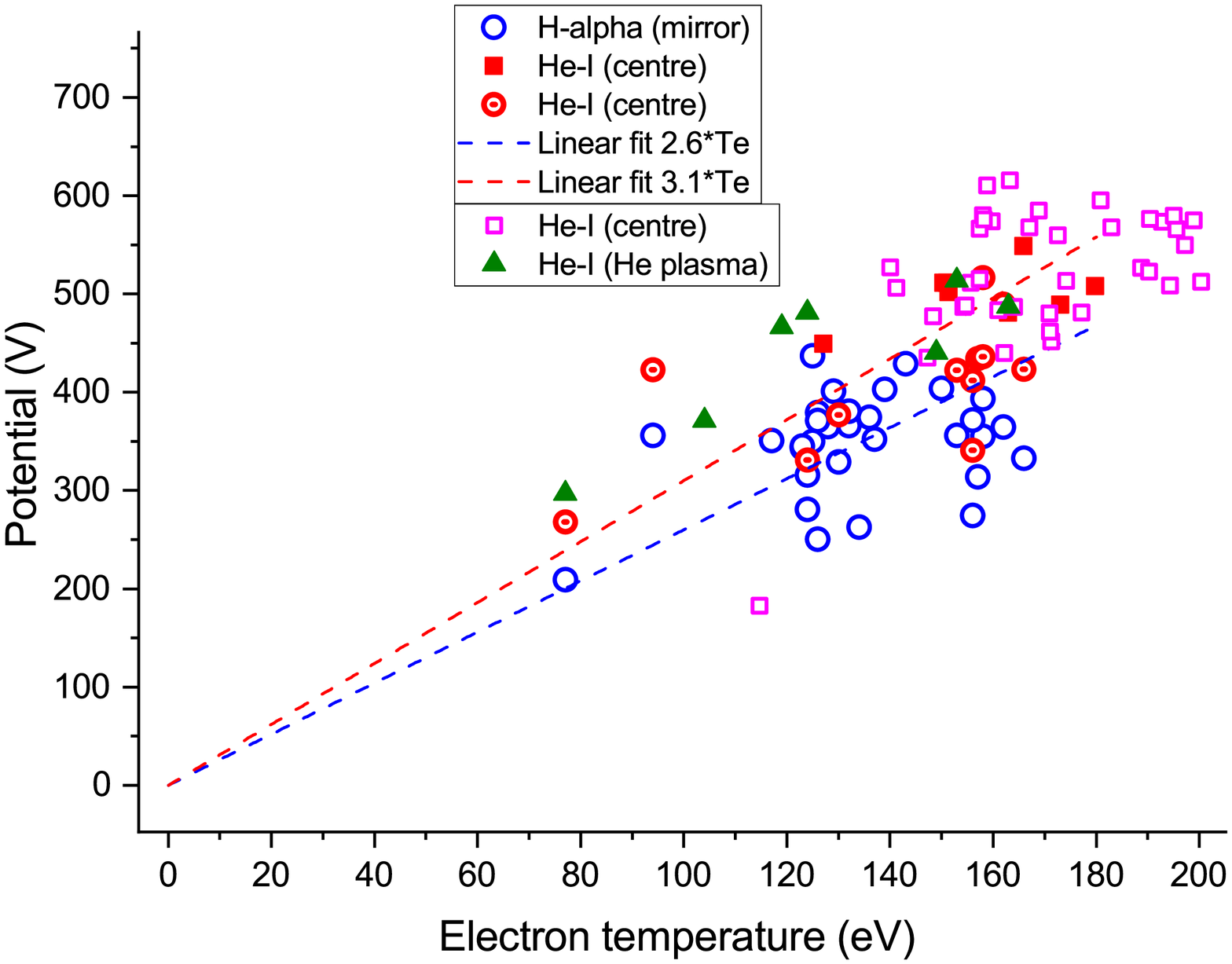}
\caption{\label{potential_Te} Ambipolar potential drop centre-expander and mirror-expander versus the on-axis electron temperature in the central plane. Blue open circles -- mirror potential calculated by the $H_{\alpha}$ Doppler shift; red filled rectangles, magenta open rectangles, red dotted circles -- centre potential calculated by the $He$-$I@667.8~nm$ Doppler shift; dark green filled triangles -- potential drop measured in the regime with $He$ target plasma and $D$ heating beams (no fit). Blue and red dashed lines show the linear fit for the mirror and the centre, respectively.}
\end{figure} 
The paper \cite{ELVIS} presents the bulk plasma ion temperature dynamics measured by the CXRS diagnostic in the same regime being discussed here. The peak ion temperature is appeared to be $\approx 15\%$ lower than the electron temperature on the axis. A separate experiment was dedicated to evaluate the on-axis ambipolar potential drop in units of the electron temperature and to calculate the relation coefficient. All numbers are measured on the axis at $t=8.5~ms$ corresponding to the peak of the fast ion energy content and $\beta_{\perp}$. Fig.~\ref{potential_Te} shows scatter plots $\Delta\varphi (T_e)$ and two linear approximations drawn for the mirror potential (blue set) and the centre potential (red and magenta sets). The statistics accumulated in this GDT regime with a moderate electron temperature, allowed to get the following coefficients: 
\begin{equation}
\label{potential_coeff}
\langle \frac{\Delta\varphi}{T_e} \rangle_{mirror} = 2.6 \pm 0.6,
\qquad
\langle \frac{\Delta\varphi}{T_e} \rangle_{centre} = 3.1 \pm 0.5.
\end{equation} 
Several shots were made with the helium target plasma and deuterium heating beams, these points are shown as filled dark green triangles in Fig.~\ref{potential_Te}. Measurements were done using the same spectral line of $He$-$I@667.8~nm$. However, in this regime we can not bind the result to the mirror region or the central region either. For the helium regime, we calculate the linear coefficient as $\langle \frac{\Delta\varphi}{T_e} \rangle_{He} = 3.5 \pm 0.4$. 

\section{Conclusion}
The pilot data acquired with the CXRS diagnostic in the GDT expander, prove that it is a robust instrument to study the physics of the plasma axial confinement even in a current status of an one-channel prototype. Making use of a gas stream CX target permitted to record optical signals with a respectable $S/N \sim 10$ on a relatively low particle density inherent to the expanding plasma flow beyond the magnetic mirror. Owing to a high magnetic expansion ratio, the gas jet does not disturb the central cell plasma parameters and provides a good spatial resolution of less than a centimetre. This experience validates important techniques of expander plasma optical diagnostics for the projects of the gas dynamic multi-mirror trap (GDMT) \cite{gdmt} and diamagnetic trap \cite{gdt_dia_trap-1} entering now the design stage.

The measured quantitative relation \ref{potential_coeff} shows the height of the ambipolar potential barrier suppressing the electron heat conductivity of the central cell plasma on the end wall. Presented numbers relate to the regime without ECRH ($T_e\leq250~eV$) and they are not far from results yielded by earlier experiments (\cite{gdt-review-ppcf}, page 27). Together with measurements of the axial energy loss per electron-ion pair \cite{stable_conf, axial_heat_loss}, the new data satisfactorily agrees with the classical model developed by Dmitri~Ryutov \cite{axial_conf}. 

The upcoming GDT experimental campaign will be supported with the upgraded CXRS diagnostic having several lines of sight to study both axial and radial profiles of the electrostatic potential. The ECRH system will be engaged as well allowing to extend the available range of electron temperatures for the study of axial plasma confinement in GDT.

\ackn
This work is supported by the Russian Science Foundation, project No.~18-72-10084 issued on 31.07.2018. 

\section*{References}
\bibliography{NF_ELVIS_lizunov}
 
\end{document}